\documentclass[conference]{IEEEtran}
\IEEEoverridecommandlockouts

\usepackage{amsmath,amssymb,amsfonts}
\usepackage{algorithmic}
\usepackage{graphicx}
\usepackage{textcomp}
\usepackage{xcolor}
\usepackage{tabularx}
\usepackage{latexsym}
\usepackage{subcaption}
\usepackage{gensymb}
\usepackage{listings} 
\usepackage{eucal}

\def\BibTeX{{\rm B\kern-.05em{\sc i\kern-.025em b}\kern-.08em
    T\kern-.1667em\lower.7ex\hbox{E}\kern-.125emX}}
\definecolor{dkgreen}{rgb}{0,0.6,0}
\definecolor{gray}{rgb}{0.5,0.5,0.5}
\definecolor{mauve}{rgb}{0.58,0,0.82}
\definecolor{lightsilver}{rgb}{0.96,0.96,0.98} 
\begin{document}

\title{Design and Simulation of 6T SRAM Array\\

}

\author{\IEEEauthorblockN{Justin London}
\IEEEauthorblockA{\textit{Dept. of Electrical Engineering} \\
\textit{University of North Dakota}\\
Grand Forks, North Dakota \\
justin.london@und.edu}

}
\maketitle

\begin{abstract}
    Conventional 6T SRAM is used in microprocessors in the cache memory design.  The basic 6T SRAM cell and a 6 bit memory array layout are designed in LEdit.  The design and analysis of key SRAM components, sense amplifiers, decoders, write drivers and precharge circuits are also provided. The pulse voltage waveforms generated for read and write operations as well as Q and Qbar nodes are simulated in LTSpice. Parasitic capacitances are extracted and their impact on the waveforms analyzed.  Static noise margin, propagation delays, and power dissipation are calculated.  Comparison of SRAM read and write operational performance using CMOS transistors is made with edge-triggered D flip flops.  If certain size area and ratio constraints are satisfied, the 6T cell with CMOS transistors will possess stability, speed, and power efficiency.  Both theoretical and simulated results are given.
\end{abstract}


\section{Introduction}
 The basic objective of a memory cell is to hold a single bit of data.  This can be achieved statically (without the need for refreshing) by using a pair of inverting gates.  Since an inverter is the smallest CMOS gate, it is ideal for use in the core of an SRAM cell.  To read from and write to this inverter pair, access transistors are required.  Though flip flops and latches can be used to store a bit, there are various problems with their design that make them unsuitable for larger memories.  For instance, D flip flops require separate data in and out lines and one cannot use a bidirectional bus.  D flip flops cannot be used a building blocks to design larger memories as a chip select input in required.  

Unlike flip-flops which are synchronous and operate on a clock signal, latches do not have a clock signal, are asynchronous, and change state continuously based on input changes.  The continuous state change can complicate static timing analysis, especially in hierarchical designs, and their user can be vulnerable to faults on the enable pin, potentially leading to unstable states, data integrity, and/or loss of data issues.  Though gated S-R latches can possibly resolve this issue, latches consume more power and area compared to flip-flops making them less ideal for applications where minimizing resource usage is essential. 

SRAM utilize pull-down (PD) resistors, pull-up (PU) resistors, and pass gate (PG) transistors. The PD and PG transistors work together in an SRAM cell to control the storage an retrieval of data.  The PD transistor ensures the output node is at the correct voltage level, while the PG transistor acts as a switch to control data flow during read and write operations. PD and PG transistors are typically NMOS transistors while PU transistors are typically PMOS. 

CMOS offer several advantages over latches including lower power consumption, higher noise margins (providing tolerance of larger variations in input voltage without errors), and greater integration density and area miniaturization. Microprocessors may contain up to 70\% of SRAMs in transistor count or area.  SRAM has two main operations: read and write.  Both operations use the same port and need sufficient noise margin.  One wordline (WL) is used access the cell.  Two bit lines (BL ("O") and BLB ("1")) are used to carry the data. Both bit lines are precharged to Vdd.  The wordline is fired for one of the cells on the bit line.  The cell pulls down either BL or BLB.  Sense amplifiers are used to regenerate the differential signal.  

SRAM cells store data as voltage levels on bit lines. Small voltage differences indicate the stored bit value. SRAMs uses row and column decoders to efficiently access specific memory locations within a 2D array structure without requiring an excessive number of input pins.  For an N-bit row address, the row decoder selects one out of $2^{N}$ word lines (rows).  The column decoder selects one out of $2^{M}$ bit lines (columns) based on an M-bit column address.   SRAM's also use sense amplifiers which are used to amplify small voltage differences between the bit lines standard logic levels (i.e. OV or Vdd)  and convert it into a robust digital signal.    The amplification process significantly speeds up the read operation through conversion of the slow, weak signal into a fast, clear digital one.  

\section{Related Works}

Various SRAM designs have been proposed from 4T to 12T.  The standard 6T design is one of the most common, due to its low leakage and compactness.  However, the design of an SRAM cell that consumes low power with very small feature size can lead to significant degradation in cell data stability as transistor properties degrade with the scaling of CMOS technology.  Moreover, SRAM cells are negatively impacted due to random variations of the doping concentration. Stability is formulated as static noise margin (SNM) defined as the maximum value of DC noise voltage that can be tolerated by the SRAM cell without altering the stored bits.  Thus, the largest noise margin is desired.  During the read operation, voltage, division between the access and driver transistors causes the read stability to be very low.

An 8T SRAM was proposed for low voltage operation by using an additional word line, increasing the mental density, wire delay and enhances capacitive coupling between adjacent word lines resulting in an increase in dynamic power consumption \cite{Chang:2008}.  A 9T SRAM cell was proposed that reduces the power by using a write bit-line balancing circuitry which increases the area overhead  \cite{Lin:2008}.  Other proposed 9T SRAM designs include one that doubles read SNM compared to conventional 6T SRAM cell but the dynamic power consumption is increased due to the conventional write operation with more transistors \cite{Liu:2008}.  Sivamangai and Gunavathi \cite{Sivamangai:2011} propose a 9T SRAM cell for high read stability and lower power consumption.  Their cell uses a single-bit line (BL) for the write operation, resulting in reduction of dynamic power consumption.  During the read operation, the data storage nodes are completed isolated from the bit lines. When a bit is read from a particular memory cell, the wordline along the cell's row is turned on high, activating all the cells in the row.  The stored value (logic 0 or 1) from the cell is sent to the bit-lines associated with it.  The bit from the selected cell is then latached fro mteh cell's amplifiier into a buffer, and put onto the output bus. 

In this project, the layout design of a 6-bit 6T SRAM cell array is proposed using LEdit.  The waveforms are simulated in LTSpice.  The parasitic capacitances are extracted in LEdit and incorporated in the circuit design in LTSpice.  The propagation delay and static noise margins are calculated with and without parasitic capacitances.  In Section III, we briefly discuss the operational methodology for SRAM read and write operations.  In Section IV, we discuss the design of the 6T SRAM cell and array including the layouts in LEdit as well as SRAM components, sense amplifiers, precharge circuits, address decoders, and write drivers.  We also discuss SRAM stability, dynamic power dissipation, static noise margin, and design using edge triggered D flip flops.  In Section V, we discuss parastic extraction and in section VI, we discuss the results with parasitic capacitances included in the cell design.  In VII, we conclude and discuss future research. Seciton VIII is an Appendix that contains the parasitic capacitance extraction for the SRAM memory array from Ledit.

\section{SRAM Operational Methodology}
In an SRAM cell, initially the bitlines are set to high voltage and the word line is set to low (BL = 1, BLB = 1, and WL = 0, assuming Q = 0 and $Q_b = 1$).  For read access, the wordlines becomes high (WL = 1).  In SRAM read operations, both bit lines (BL and BLB) are precharged to Vdd.  The wordline WL is activated (fired) for one the cells on the bit line.  The cell pulls down either BL or BLB.   Data should not flip after the read access.  

In SRAM designs, the driving transistors (known as write drivers) used for writing data to the memory cells must be stronger than the access transistors used for reading data. This is crucial for ensuring accurate and reliable write operations and is also essential for maintaining the stored data in the cell.  The bitline delay $t^{B}_{delay} = \frac{C_{B}\Delta V_{B}}{I_{cell}}$  The bitline capacitance $C_{B}$ is large due to the large number of cells attached.  The cell current $I_{cell}$ is small due to high density cells.  $\Delta V$ has to be minimized for high speed.  

When a cell is not accessed such that the worldine is low (WL = 0), data is safely held inside the cell and the noise margin is high.  When a cell is accessed and  the wordline is high (WL= Vdd), the access transistor acts as a noise source.  This may lead to to data '0' pulled up to the cell's data line voltage high, and potentially causing the cell data to flip if the threshold voltage is overcome.  The flipping of cell data during a read operation can can lead to read failure.  To improve read noise margins, one can boost the cell voltage supply.  This makes the driver stronger than acess, suppressing the rise in the low side and effectively improves the beta ratio.  The NMOS driver can also be downsized decreasing the cell size.

\section{Design}

\subsection{Cell Layout}
\begin{figure}[h]
\centering
	\begin{center} 
	\includegraphics[width=0.7\linewidth]{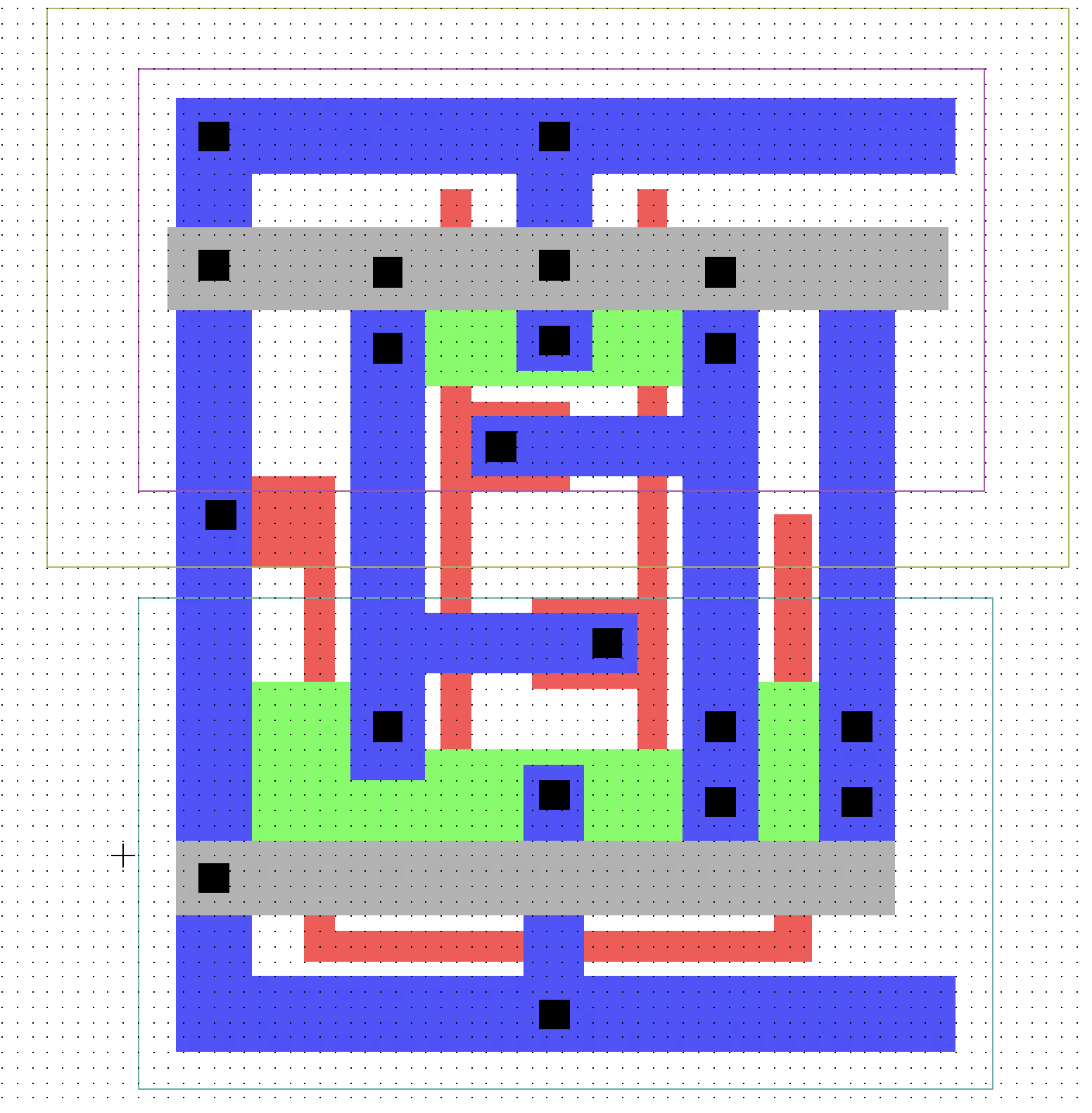}
	\caption{6T SRAM Cell} 
	\label{fig:cell} 
	\end{center} 
\end{figure}

\begin{figure}[h]
\centering
	\begin{center} 
	\includegraphics[width=1\linewidth]{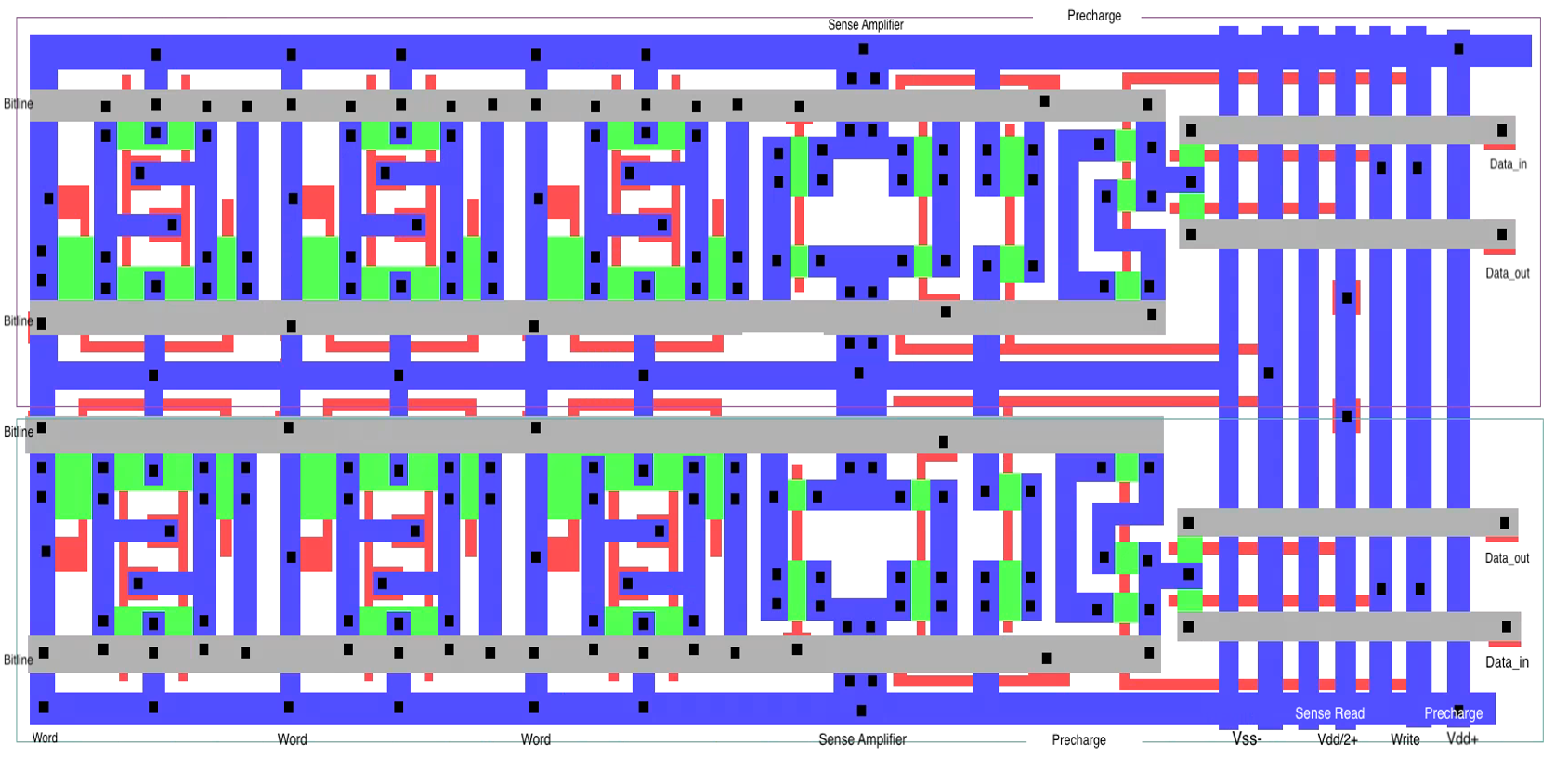}
	\caption{6 Bit SRAM Array} 
	\label{fig:cell2} 
	\end{center}  
\end{figure}

\begin{figure}[h]
\centering
	\begin{center} 
	\includegraphics[width=0.9\linewidth]{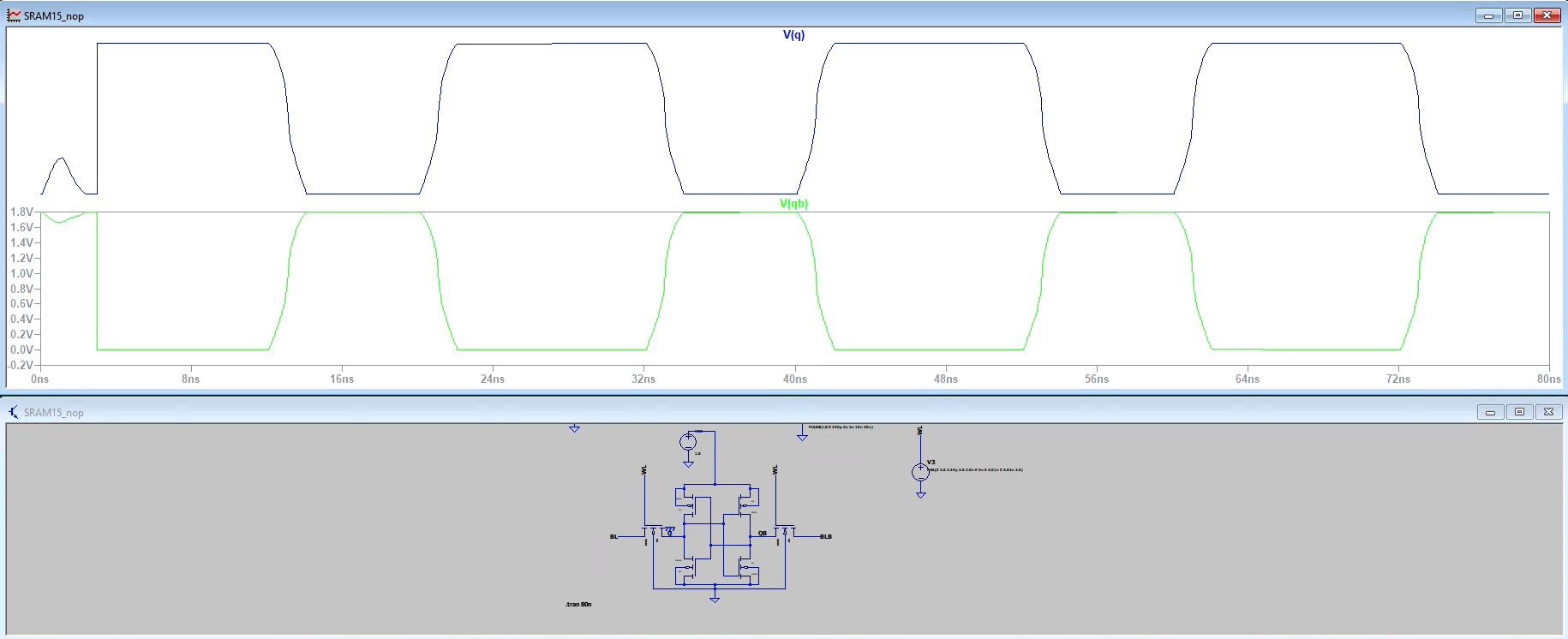}
	\caption{Waveforms of Q and Q\_bar Nodes} 
	\label{fig:volt} 
	\end{center}  
\end{figure}

    Figure \ref{fig:cell} shows a layout of the 6T SRAM cell  in LEdit.  Two Metals, M1 and M2 are used for the wordlines and bitlines along with polysilcon for routing.   The areas of the PMOS is $1,568 \lambda m^{2}$, the NMOS is $1,792 \mu m^{2}$, and the total cell area is $4,364 \mu m^{2}$.   However, when the NWell is added around the PMOS, the cell size gets larger.  The PMOS width W = 67.5 $\lambda$, height H = $37 \lambda$, and the area is 2497.50 $\lambda m^2$. The NMOS W = $56.5 \mu$, $H = 32.5 \mu$, and area = $1836.25 \lambda m^2$.  The total cell area is $4,446.75 \lambda m^2$.
    A layout of the 6 Bit SRAM Array is shown in Figure \ref{fig:cell2}.    The total area of the array is 41,162.3 $\lambda m^2$ where the PMOS is 18,676 
    $\lambda m^2$ and 22,477.50 $\lambda m^2$.
    The 6T SRAM cell circuit design in LTSpice is shown in Figure \ref{fig:sense}.  The sense amplifier in the SRAM cell (at the end of two complementary bit-lines) detect and amplify low power signals from a bit line that represents a data bit stored in the memory cell, and amplify small voltage swings to nominal logic levels.  
\begin{figure}[h]
\centering
	\begin{center} 
	\includegraphics[width=0.9\linewidth]{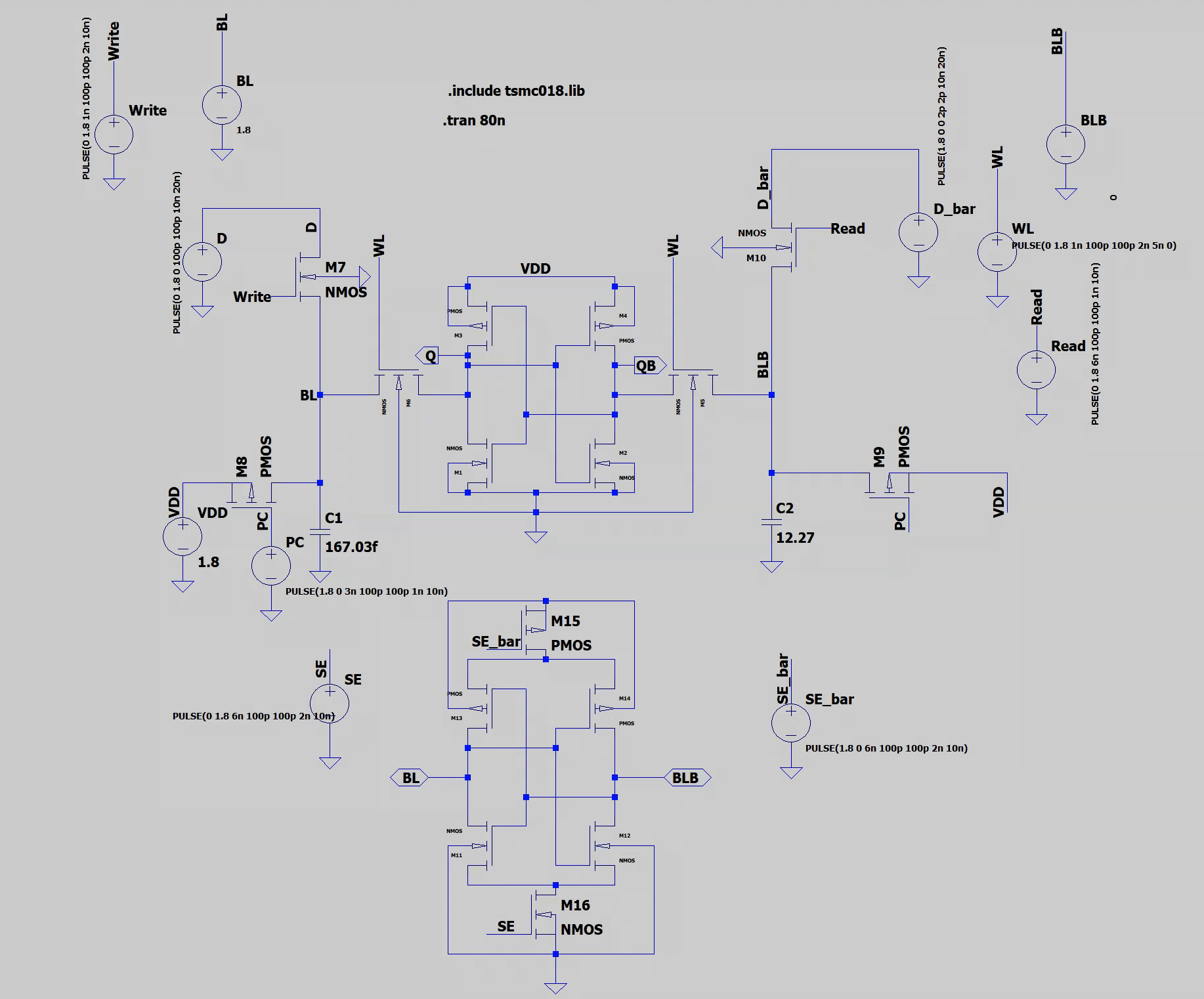}
	\caption{6T SRAM Cell Circuit} 
	\label{fig:sense} 
	\end{center}  
\end{figure} 
    The voltage waveform at Q and Qbar nodes are shown in Figure \ref{fig:volt}.  The propagation delay is calculated as
    \begin{equation}
        t_{p} = \frac{t_{pLH} + t_{pHL}}{2} = \frac{12.01 ns + 12.15 ns}{2} = 12.08 ns.
    \end{equation}
    
\begin{figure}
\centering
	\begin{center} 
	\includegraphics[width=0.9\linewidth]{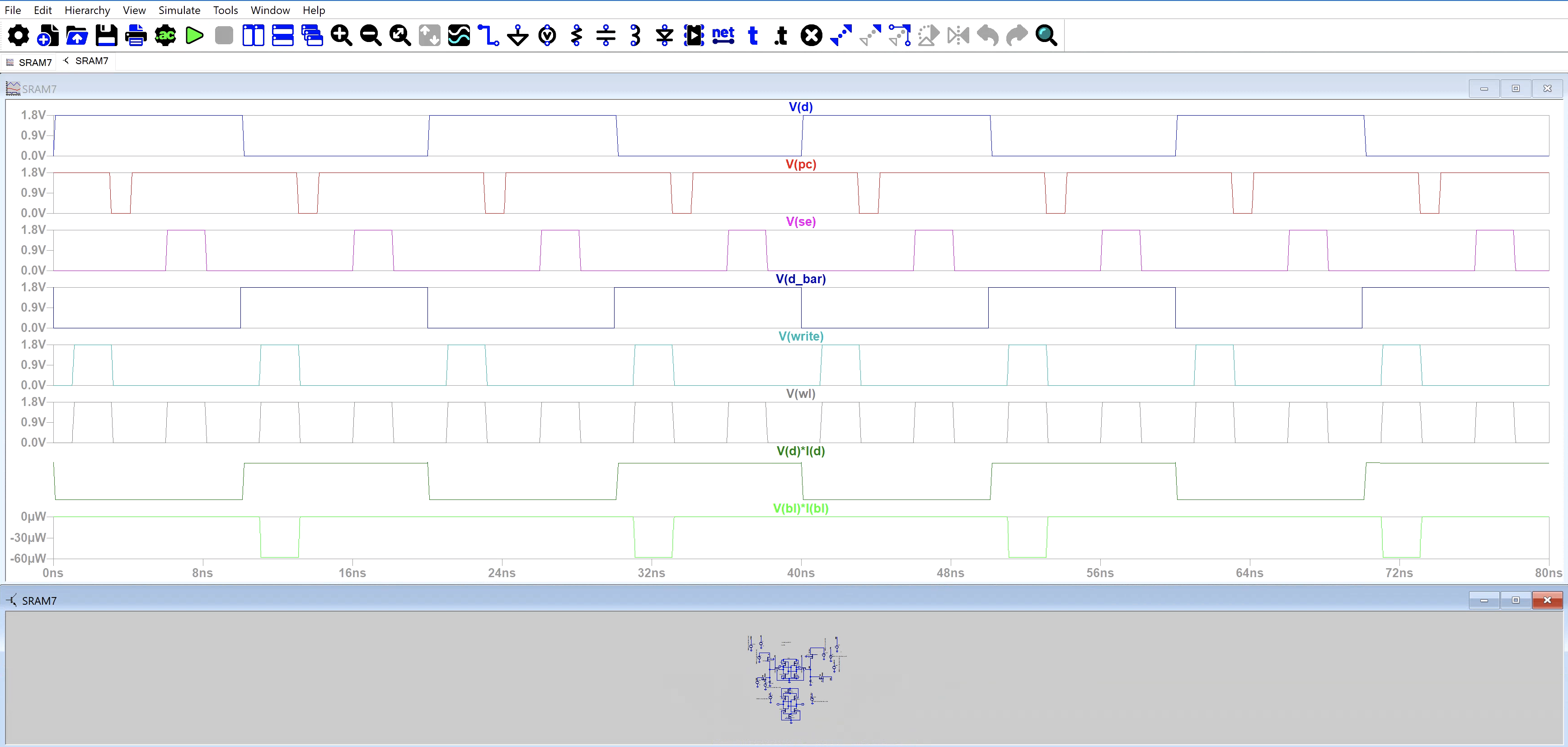}
	\caption{Waveforms for 6T SRAM Cell} 
	\label{fig:cellwaveform} 
	\end{center}  
\end{figure}

    The voltage and power waveforms at different nodes are  shown in Figure \ref{fig:cellwaveform}
    The voltage pulse waveforms for the dataline (d), precharge (pc), sense amplifier (se), d\_bar, write, wordline (WL) nodes and the power of the dataline  and bitlines of the 6T SRAM cell are shown in Figure \ref{fig:cellwaveform}.  Vdd = 1.8 V.

\subsection{Sense Amplifier and Precharge}
The sense amplifier used in DRAM memory circuits are similar to those for SRAM except they perform an additional function, memory refresh:  The data in DRAM chips is stored as electric charge in tiny capacitors in the memory cells. The read operation depletes the charge in a cell, destroying the data, so after the data is read out of the cell, the sense amplifier must immediately write it back in the cell by applying a voltage to it, recharging the capacitor.  
The sense amplifier amplifies a small analog differential voltage developed on the bit lines in a read acess.  The amplification results in full swing single digital output.  Sense amplifiers reduce the size of the SRAM cell since the driver transistors do not need to fully discharge the bitlines.  Typically, read operations are the slowest which gives delay in the cell. the read margin is the bit line 
differential required to activate the sense amplifier during a read operation, ensuring accurate data retrieval.  Bit lines have larger capacitances due to the metal length and number of transistors that will consume greater amount of  time to discharge the bit lines.    

The SRAM also uses a precharge circuit for the read and write operations for precharging and equalizing the bit-line (BL)O and bit-line bar (BLB) lines.  Figure \ref{fig:sense} shows an LEdit layout of the sense amplifier and precharge and their corresponding LTSpice circuits.  The sense amplifier uses 3 PMOS and 3 NMOS transistors.   The precharge circuit uses 3 PMOS transistors.
\begin{figure}[h]
\centering
	\begin{center} 
	\includegraphics[width=0.9\linewidth]{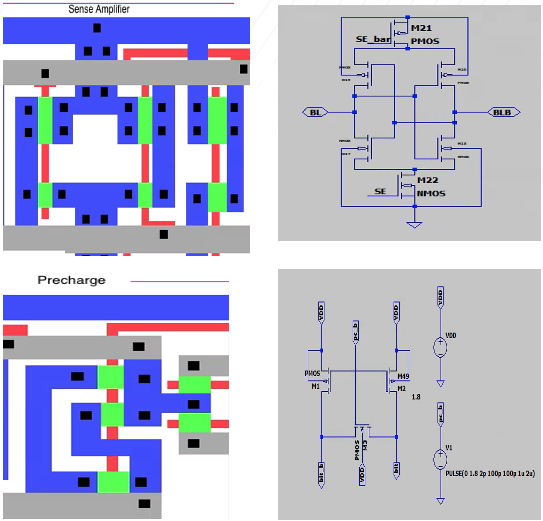}
	\caption{Sense amplifier (top) and precharge circuit (bottom)} 
	\label{fig:sense} 
	\end{center}  
\end{figure} 

\subsection{Address Decoder and Write Driver}
\begin{figure}[h]
\centering
	\begin{center} 
	\includegraphics[width=0.9\linewidth]{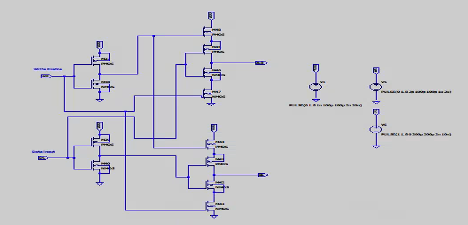}
	\caption{Write Driver Circuit in LTSpice} 
	\label{fig:writedriver} 
	\end{center}  
\end{figure}

\begin{figure}[h]
\centering
	\begin{center} 
	\includegraphics[width=0.9\linewidth]{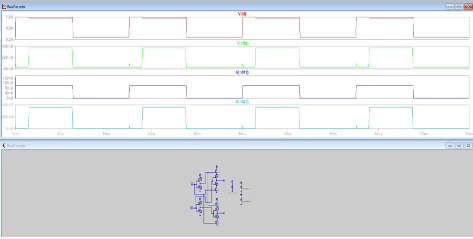}
	\caption{Write Driver Waveforms} 
	\label{fig:driver2} 
	\end{center}  
\end{figure}

    The address decoder is used to decode the input address and enable the word line.   A 2:4 NAND CMOS decoder attached to the cross-coupled inverter for access in the write phase.  Two dynamic NAND CMOS decoders are used.  One is used as a row decoder that selects a particular word line by raising the voltage level.  The second is a column decoder selects the particular bit line column.
\begin{figure}[h]
\centering
	\begin{center} 
	\includegraphics[width=0.8\linewidth]{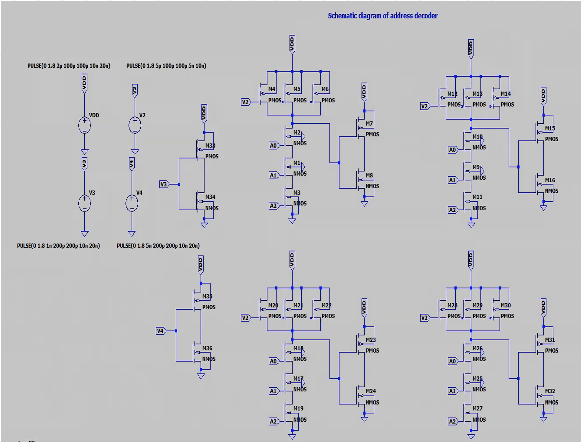}
	\caption{Address Decoder} 
	\label{fig:decoder} 
	\end{center}  
\end{figure} 
    The write driver has the duty of discharging the bit lines to a level below the write margin of the cell quickly before or while the word lines of the selected cell are active.  The write margin is the minimum bit line voltage needed to flip the state of an SRAM cell during a write operation, ensuring successful data storage.  Two typical write drivers in LTSpice are shown in Figure \ref{fig:writedriver}.  The data input selects which bit line is discharged.  The Word Enable (WE) is turned on only when the write operation is intended.  Otherwise, the WE isolates the bit lines from the write drivers. The write driver is made up of a CMOS inverter and NMPOS pass transistor and it does not depend on the number of transistors to form a column. 
    Waveforms generated by the write driver are shown in Figure \ref{fig:driver2}.

\subsection{Stability}
    Cell ratio (CR) values (W/L) should be greater than one to ensure a stable read operation.   The CR should satisfy:
\begin{equation}
    CR = \frac{(W/L)_{PD1}}{(W/L)_{PG1}} = \frac{(W/L)_{PD2}}{(W/L)_{PG2}}
    \label{eq:cr}
\end{equation}
    Writability imposes a constraint where the PG transistor must be stronger than the corresponding PU transistor to copy the BL value into the SRAM cell \cite{Gul:2022}.   The pull ratio (PR) should be less than one for the write operation to be stable:
\begin{equation}
    PR = \frac{(W/L)_{PU1}}{(W/L)_{PG1}} = \frac{(W/L)_{PU2}}{(W/L)_{PG2}}
    \label{eq:pr}
\end{equation}
    To analyze the stability, we use the extracted PMOS and NMOS measurements provided from LEdit in Section V.
    The proposed 6T cell design only satisfies the LHS  of (\ref{eq:cr}) of the CR ratio read constraint of [(10.5/2)/(10.5/2.5) = 1.25] with widths of $10.5 \mu$ and lengths of $2 \mu$ for PU1 and PU2, respectively.  The widths and lengths for both PG1 and PG2 are 10.5 $\mu$ and $2.5 \mu$, respectively.   The width and length of PD1 and PD2 are $6 \mu$ and $2 \mu$, respectively.  The RHS ratio in (\ref{eq:cr}) is (6/2)/(10.5/2.5) = 0.714, which is less than 1 and does not equal the LHS ratio.  While the design satisfies the PR constraint as the RHS = LHS in (\ref{eq:pr}), it does not satisfy the writability constraint as the ratios are 1.25 which is greater than 1.  

\subsection{Dynamic Power Dissipation}

\ \ \ Dynamic power dissipation occurs due to charging and discharging of load capacitance and short circuit current when both the NMOS and PMOS devices remain ON for short time durations \cite{Sivamangai:2011}.    Load capacitance is typically the dominant term.  The average dynamic power ($P_{d}$) dissipated is given by:
\begin{align}
    P_{d} &= C_{L}V_{DD}^{2}f_{sw} 
\end{align}
Assuming $C_{L} = 35 fF, V_{DD} = 1.8 V$, and $f_{sw} = 100MHz$, then
\begin{align}
P_{d} &=  (35 fF)(1.8 V)^2(100 MHz) \\
          &= 6.3 \times 10^{-6} \ W = 0.0063 \ mW
\end{align}
The power dissipated is proportional to the charging and discharging of load capacitance $C_{L}$, supply voltage ($V_{DD}$) and the switching frequency $f_{sw}$.  A reduction in the voltage supply can lead to degradation of the cell data stability \cite{Mohagheghi:2022}.  

\subsection{Static Noise Margin}

\ \ \ The transistors must satisfy ratio constraints to ensure both read and write stability.   The NMOS pulldown transistor in the cross coupled inverters must be strongest, the access transistors are of intermediate strength, and the pMOS pullup transistors must be weak.  A cell's stability and writability are quantified by the hold margin, the read margin, and the write margin, which are determined by the static noise margin of the cell in its various modes of operation.  A cell should have two stable states during hold and read operations and only one stable state during write.  

The static noise margin (SNM) measures how much noise can be applied to the inputs of the two cross-coupled inverters before a stable state is lost (during hold or read) or a second stable state is created during write. The hold margin is the SNM while the cell is holding its state and being neither read or written.  A noise source $V_{n}$ is applied to each of the cross-coupled inverters.          
The static noise margin (SNM) of the cell can be determined graphically from the butterfly diagram.  The plot is generated by setting the noise source $V_{n} = 0$ and plotting $V_{1}$ and $V_{2}$, the voltage for each of the inverters.  If the inverters are identical, the DC transfer curves are symmetric and mirrored across the 45 degree line of $V_{1} = V_{2}$ and the high and low SNM are equal.  If the inverters are not identical, the SNM is the lesser of the two.  The SNM increases with $V_{DD}$ and threshold voltage $V_{th}$. 

\begin{figure}[h]
\centering
	\begin{center} 
	\includegraphics[width=0.9\linewidth]{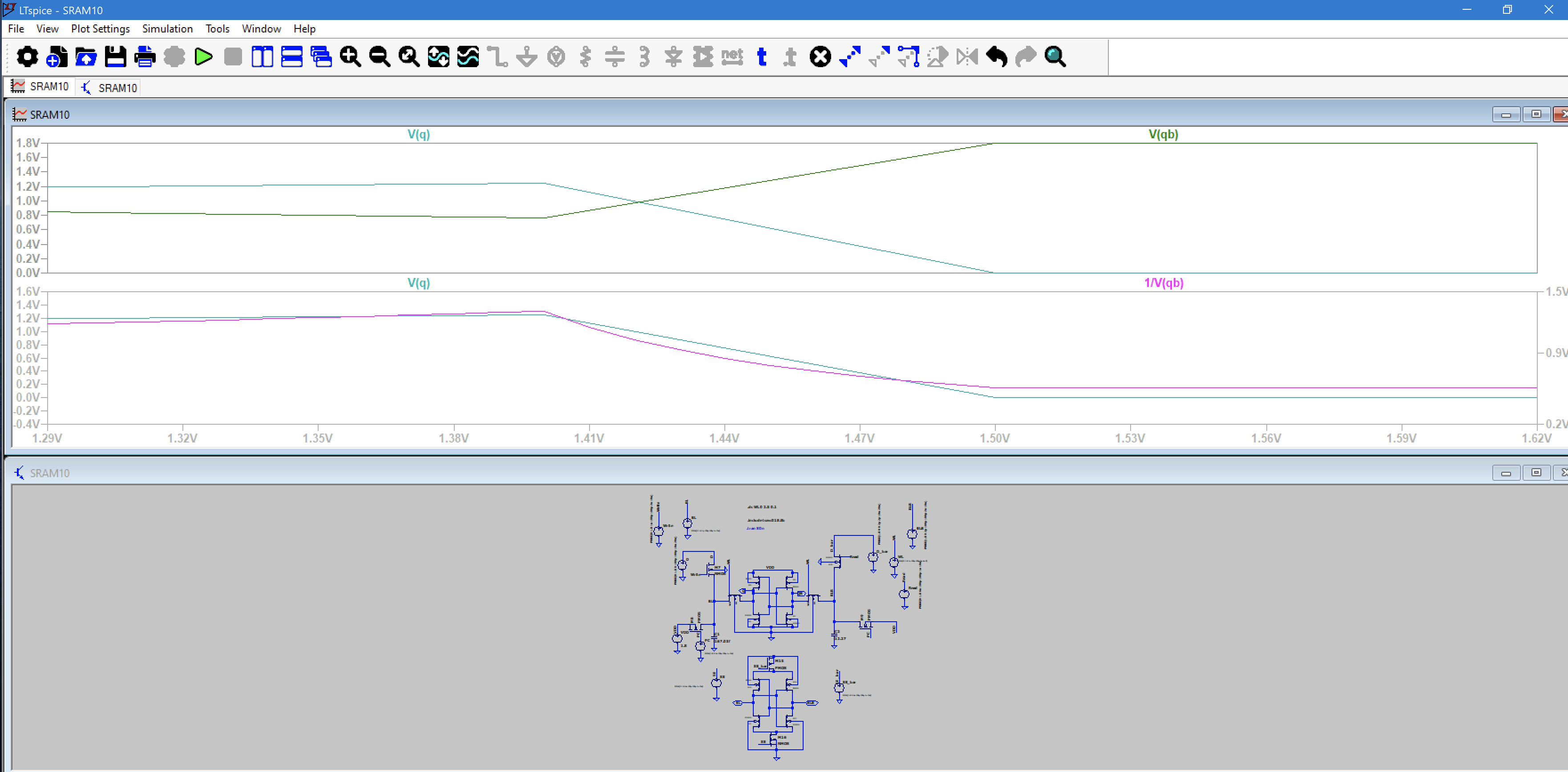}
	\caption{Butterfly Curves} 
	\label{fig:butterfly} 
	\end{center}  
\end{figure}

The butterfly plot shows two stable states (with one input low and the other high) and one metastable state with $V_{1} = V_{2}$. Excessive noise eliminates the stable state of $V_{1} = 0$ and $V_{2} = V_{DD}$, forcing the cell into the opposite state.  The SNM is determined by the length of the side of the largest square that can be inscribed between the curves.  Figure \ref{fig:butterfly} shows a plot of the butterfly curves.
    The SNM is 3.09 mV.  Aside from computing the maximum square between the normal and mirrored VTC, other methods can be used to calculate SNM including small-signal loop-gain approach, Jacobian of the Kirchoff equations, and coinciding roots.  It has been researched and shown that these methods are all equivalent.   Qin \cite{Qin:2000} adopts a linear macro-model given by 
    \begin{equation}
        SNM = k(V_{dd} - DRV)
    \end{equation}
    where $k \approx 2/(3+n)$ and DRV is the data retention voltage and a measure of the SRAM cell's state-retention capability under very low voltage and $n$ is the noise tolerance factor.  In an ideal CMOS, $n=1$.  In order to reliably preserve data in an SRAM cell, the cross-coupled inverters must have a loop gain greater than one.  However, when $V_{DD}$ scales down to DRV, the VTC of cross-coupled inverters degrade to a level such that the loop gain reduces to one and the SNM of the RAM cell falls to zero \cite{Qin:2000}  
        During standby operation mode, the bitline voltages are connected to Vdd and reduced to DRV.  All six transistors in the SRAM are in the sub-threshold region and $n_{i}$ are the sub-threshold factor (sub-threshold swing divided by 60mV at room temperature.)  
        The threshold voltage is the minimum gate voltage required to switch the device between on and off states.  The threshold voltage of the i$th$ transistor is:
        \begin{multline}
            V_{th,i}=V_{th,0} + \gamma_{i}\bigg (\sqrt{\lvert -2 \phi_{F.i} + V_{SB,i} \rvert} - \sqrt{\lvert -2\phi_{F.i} \rvert} \bigg ) \\ -  V_{DS,i}\text{exp}({-\alpha L_{i})}
        \end{multline}  
    where the second and third terms represent the body bias and the drain-induced-barrier-lowering (DIBL) effect, respectively. \cite{Yu:1997}.  $v_{th,0} = \phi_{ms} - 2\phi_{F,i} - \frac{Q_{B0} + Q_{ox} + Q_{I}}{C_{ox}}$; $\phi_{i}$ is the surface potential at strong inversion, representing the voltage required to create a conductive MOS channel.  The body-effect coefficient $\gamma = \frac{\sqrt{2q\epsilon_{si}N_{A}}}{C_{ox}}$ with the oxide permittivity $\epsilon_{ox} = 3.97 \times \epsilon_{0} = 3.5 \times 10^{-13} F/cm$ and $C_{ox} = \frac{\epsilon_{ox}}{t_{ox}}$ with the oxide thickness $t_{ox} = 20 nm$ or smaller \cite{Rabaey:1996}. 
    
        The standard deviation $\sigma_{V_{th}}$ of the threshold voltage depends on transistor length and width.  Let $A_{V_{th}}$ be technology-dependent parameter.  $V_{th}$ variation increases with transistor dimension reduction since process variations increase.  Experiments have shown that in general, SNM under process, voltage, and temperature (PVT) variations is roughly distributed as a Gaussian distribution with a mean value $\mu_{SNM}$ and $\sigma_{SNM}$ \cite{Hamouche:2012}.  Moreover, 
            \begin{equation}
                \sigma_{V_{th}} = A_{V_{th}}\sqrt{\frac{1}{WL}}
            \end{equation}
        Let $I_{i}$ be the sub-threshold current of the $i$th transistor. $I_{i}$ can be modeled as:
        \begin{equation}
            I_{i} = I_{off,i}\text{exp} \bigg (\frac{V_{GS,i}}{n_{i}v_{T}} \bigg )\bigg (1 - \text{exp}\bigg (\frac{-V_{DS,i}}{v_{T}} \bigg ) \bigg  )  
        \end{equation}
            where 
        \begin{equation}
            I_{off,i} = \beta_{i}I_{0}\text{exp} \bigg (\frac{-V_{th,i}}{n_{i}v_{T}} \bigg ) 
        \end{equation}
    and $v_{T} = \frac{kT}{q}$ is the thermal voltage equal to 26 mV when $T = 27^{\circ} C$; $\beta_{i}$ is the transistor ($W/L$) ratio; $I_{0}$ is the leakage current of unit sized device at $V_{GS} = 0$ and $V_{ds} >> v_{T}$; and $T$ is the chip temperature.
        DRV can be calculated as:
    \begin{equation}
        DRV = DRV^{(0)} + \bigg [\frac{V_{1}}{2} + \frac{(DRV^{(0)} - V_{2})n_{2}}{2} \bigg ]
    \end{equation}
    where 
    \begin{multline}
        DRV^{(0)} = \frac{kT/q}{n^{-1}_{2} + n^{-1}_{3}}\text{log}\bigg [(n^{-1}_{3} + n^{-1}_{4})\frac{I_{off,4}}{I_{off,2}I_{off,3}} \cdot \\ 
        \bigg (\frac{I_{off,5}}{n_{2}} + \frac{I_{off,1}}{(n^{-1}_{1} + n^{-1}_{2})^{-1}}\bigg ) \bigg ]
    \end{multline}
    and 
    \begin{align}
        V_{1} &= \frac{kT}{q}\cdot \frac{I_{off,1} + I_{off,5}}{I_{off,2}}\text{exp}\bigg (\frac{-DRV^{(0)}}{n_{2}kT/q}  \bigg ) \\
        V_{2} &= DRV^{(0)} -\frac{kT}{q}\cdot \frac{I_{off,4}}{I_{off,3}}\text{exp}\bigg (\frac{-DRV^{(0)}}{n_{3}kT/q}  \bigg )
    \end{align}
    The formula for DRV only relies on the values of $I_{off,i}$ and $n_{i}$ which can be extracted from transistor properties using simulation or measurement \cite{Qin:2000}.  For 130nm industrial technology, n=1.25 for both PMOS and NMOS.  In  ideal CMOS technology (n=1), $DRV_{ideal} = 2v_{T}\text{ln}(1+n) = 36mV$.
\subsection{16-Bit Edge Triggered SRAM Design}
\begin{figure}[h]
    \centering
	\begin{center} 
	\includegraphics[width=0.9\linewidth]{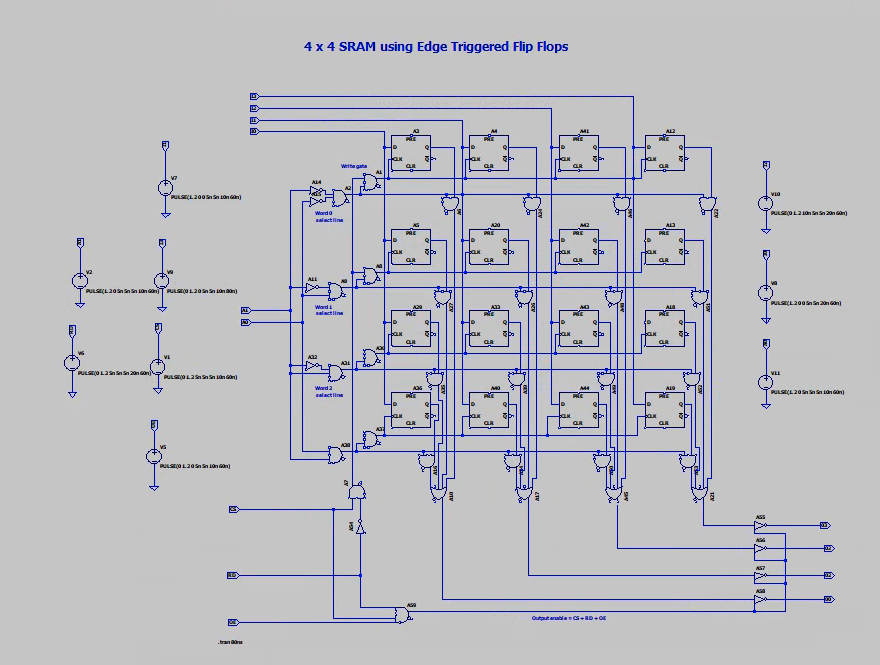}
	\caption{$4 \times 4$ SRAM 4 Edge Triggered Flip Flops} 
	\label{fig:edge7} 
	\end{center}  
\end{figure}

    A $4 \times 4$ SRAM using edge triggered D flip flops as designed in LTSpice as shown in Figure \ref{fig:edge7}.  Unlike CMOS transistors, edge-triggered flip-flops change state (store a new bit) only when the clock signal experiences a rising or falling edge.  Thus, data is stored stability with respect to the clock signal.   

    When a write operation occurs, the flip-flop within the SRAM cell is set to the desired state (0 or 1). The flip-flop holds this state until a new write operation or power is removed. During a read operation, the flip-flop's state is retrieved, allowing access to the stored data.
    
    The clock signal triggers the flip-flop to update its output based on the current input state.  The output enable input controls whether the flip-flop's output is active or disabled.  When OE is enabled (usually 1), the flip-flop's output can change. When OE is disabled (usually O), the output is held in its current state, regardless of clock transitions or other inputs.  The reset input forces the flip-flop's output to a specific state (usually O). When the reset signal is activated, the flip-flop's output is immediately reset, regardless of its current state or the clock.

    To prevent flip flops changing to the wrong state, one must be consider worst-case static noise, defined as DC disturbance which is adversely present in all logic gates chained to each other in exactly the same way \cite{Lohstroh:1983}.  The network equation in terms of any node voltage of a flip flop will be cubic for binary logic, fifth degree for ternary logic so that there will be at most three, five, etc. real roots, respectively.  One must check that small loop gain is not unity which will cause a stable state to become metastable.  This is equivalent the Jacobian of the Kirchoff equations being 0.

\begin{figure}[h]
\centering
	\begin{center} 
	\includegraphics[width=0.8\linewidth]{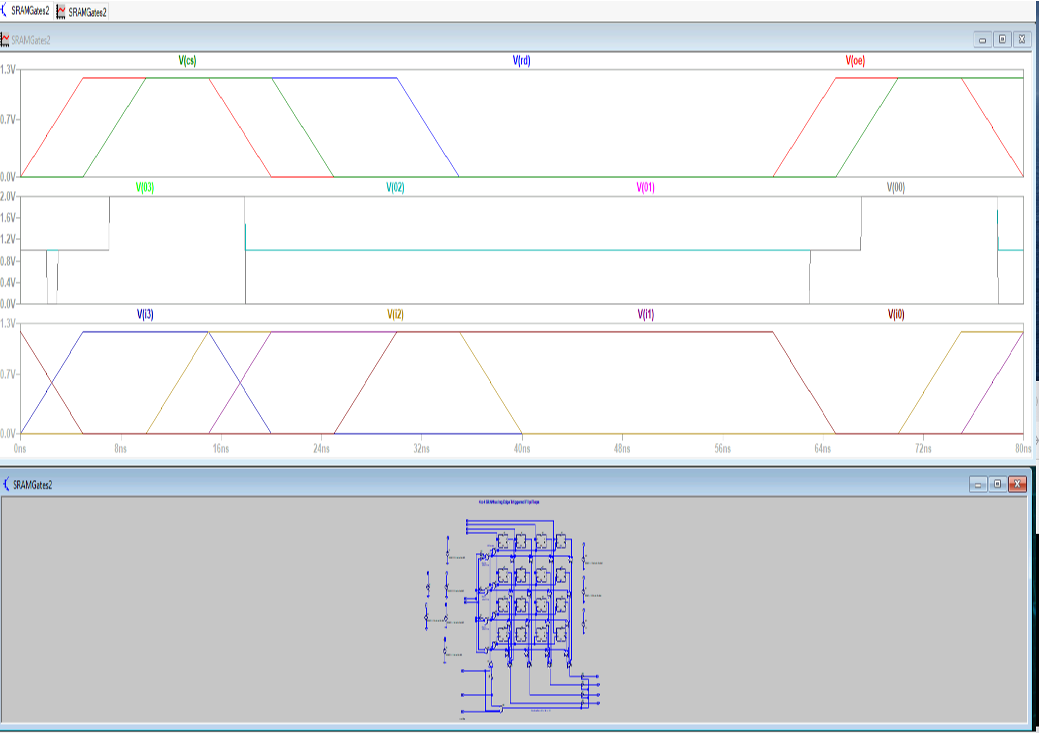}
	\caption{} 
	\label{fig:edge2} 
	\end{center}  
\end{figure}

\section{Parasitic Extraction}
\begin{figure}[h]
    \centering
	\begin{center} 
	\includegraphics[width=0.8\linewidth]{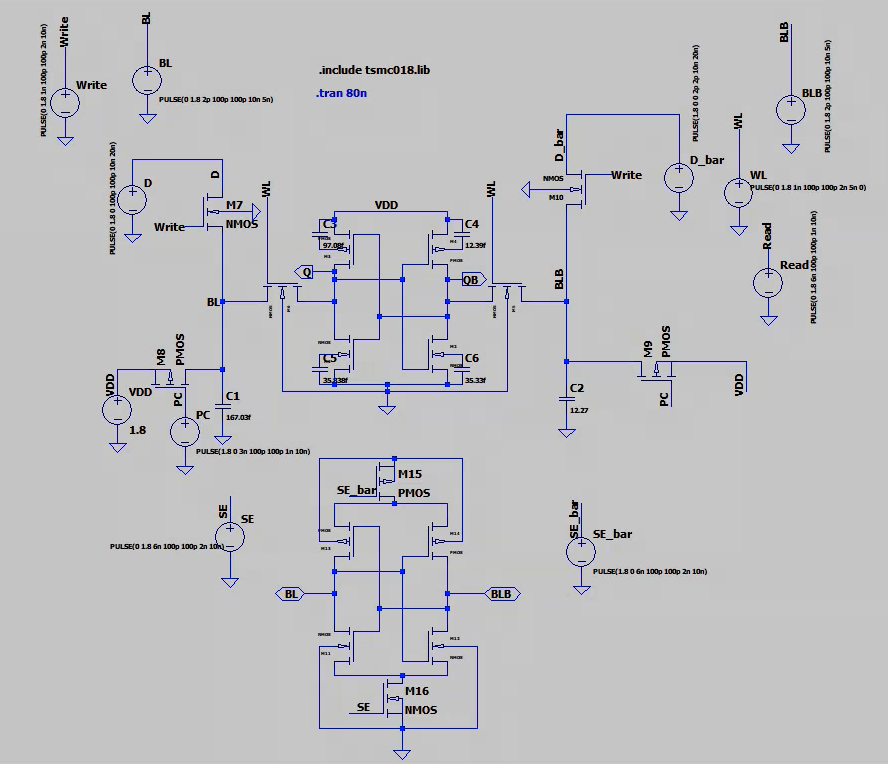}
	\caption{6T SRAM with Parastic Capacitances} 
	\label{fig:ltspice} 
	\end{center}  
\end{figure}

\begin{figure}[h]
    \centering
	\begin{center} 
	\includegraphics[width=0.8\linewidth]{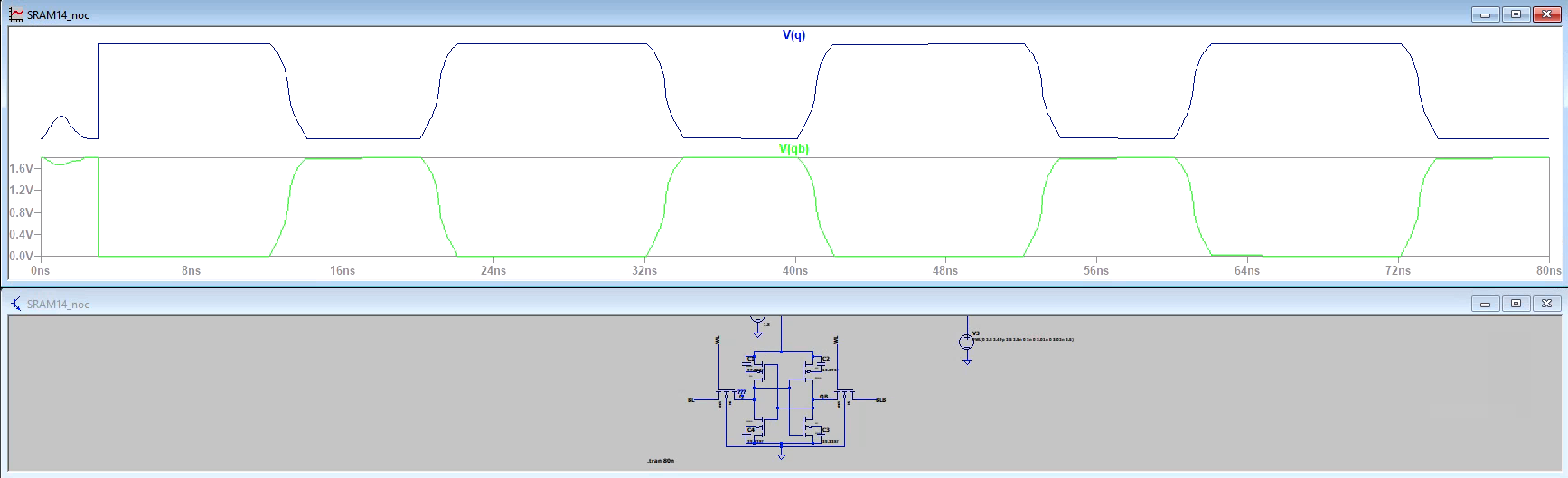}
	\caption{Voltage Waveforms for Q and Qbar with parasitics} 
	\label{fig:pc} 
	\end{center}  
\end{figure}

\begin{figure}[h]
    \centering
	\begin{center} 
	\includegraphics[width=0.8\linewidth]{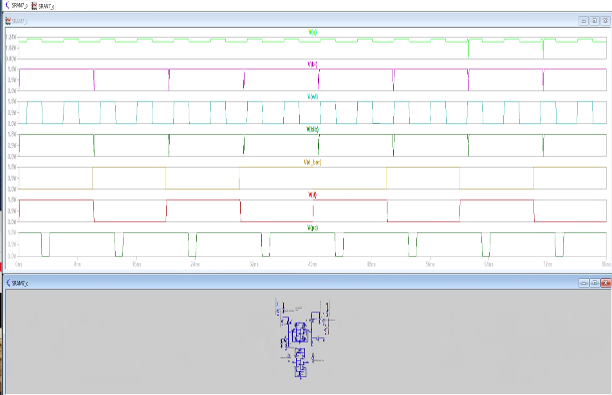}
	\caption{Waveforms with Parastic Capacitances} 
	\label{fig:cap} 
	\end{center}  
\end{figure}

    Extracting the parasitic capacitances for the single SRAM cell in LEdit yields:
\begin{lstlisting}
Cpar1 1 0 C=97.083f
Cpar2 2 0 C=12.392f
Cpar3 3 0 C=35.838f
Cpar4 4 0 C=35.338f
* WARNING: Node 5 has zero nodal parasitic capacitance.
* WARNING: Node 6 has zero nodal parasitic capacitance.

M5 3 4 1 6 PMOS L=2u W=10.5u AD=63p PD=33u AS=143p PS=64u 
* M5 DRAIN GATE SOURCE BULK (34 31 36 41.5) 
M6 1 3 4 6 PMOS L=2u W=10.5u AD=143p PD=64u AS=63p PS=33u 
* M6 DRAIN GATE SOURCE BULK (21 31 23 41.5) 
M7 3 4 1 5 NMOS L=2u W=6u AD=71.5p PD=36u AS=222.75p PS=109u 
* M7 DRAIN GATE SOURCE BULK (34 1 36 7) 
M8 1 3 4 5 NMOS L=2u W=6u AD=222.75p PD=109u AS=71.5p PS=36u 
* M8 DRAIN GATE SOURCE BULK (21 1 23 7) 
M9 2 1 3 5 NMOS L=2.5u W=10.5u AD=57.75p PD=32u AS=71.5p PS=36u 
* M9 DRAIN GATE SOURCE BULK (43 1 45.5 11.5) 
M10 4 1 1 5 NMOS L=2u W=10.5u AD=71.5p PD=36u AS=222.75p PS=109u 
* M10 DRAIN GATE SOURCE BULK (12 1 14 11.5) 

* Total Nodes: 6
* Total Elements: 10
* Extract Elapsed Time: 0 seconds
.END
\end{lstlisting}
    Using these parasitics in the LTSpice circuit, the modified circuit in LTSpice is shown in Figure \ref{fig:ltspice}
    The voltage waveforms for Q and Qbar are shown in Figure \ref{fig:pc}.
    The propagation delay is calculated as
    \begin{equation}
        t_{p} = \frac{t_{pLH} + t_{pHL}}{2} = \frac{11.89 ns + 12.09 ns}{2} = 11.99 ns \approx 12ns.
    \end{equation}
    Thus, the parasitic capacitances increase the propagation delay for the SRAM cell by approximately 7 ns compared to the cell without parasitic capacitances.  In large SRAM memory arrays with many logic gates and cells and/or long wordlines and bitlines, the propagation delay will accumulate and be much greater.
    The waveforms of this circuit are shown in Figure \ref{fig:cap}.   The voltage waveforms of the BL and BLB no longer have smooth transitions from VDD to VSS (GND), but are sharp.  
\section{Results}
    As the parastic capacitance results show, parasitics can introduce read and write delays and possibly failure.  The power dissipation of SRAM decreases as the cell size is reduced as the size of the parasitic (load) capacitance is reduced.  However, the read and write margins (namely, the conditions udner which the cell can reliability retain and read data) generally decreases.  This is primarily due to several factors, including increased susceptibility to noise interference, leakage currents, and process variations.  Since the signal voltages are smaller in smaller cells, the transistors are more sensitive to electrical noise variations in operating conditions.  The leakage currents can drain the bit-line voltage during read and write operations, making it harder to maintain the correct logic levels and reducing the cell's stability.  Moreover, manufacturing imperfections and variations in transistor properties such as threshold voltages, on-state currents) become more pronounced as cell size shrinks causing different cells within the same SRAM array to have different read and write margins. 

\section{Conclusion}
    A 6T SRAM cell layout and array design was proposed.   6T SRAM are the most widely used memory cells due to their compact size and efficiency.  The W/L ratio of the transistors in SRAM cell impact the stability.  They are quite efficient with high resistance to voltage variation and transient noise.  CMOS transistors are less dense and more area-efficient than edge triggered flip flops in design and therefore should be used for large scale memory arrays.  The SNM can be generated from the  butterfly curves generated by the voltage transfer characteristic (VTC) curves of the cross-coupled inverters.  The sense amplifiers and precharge circuitry provides important capabilities for read access and speed in contrast to the use of floating charge capacitors used in DRAM.  Future research will involve designing SRAM cells and arrays with other topologies and DRAM arrays, and comparing their operational characteristics, efficiency, and stability with 6T SRAM. The widths and lengths of the PMOS and NMOS need to be modified to ensure both the CR and PR constraints are satisfied.  
    
\section{Appendix}
    The parastic capacitances of the SRAM Array extracted in LEdit from the layout in Figure \ref{fig:cell2} are shown below.  However, there were some challenges in the SRAM array design.  The PSelect and NSelect could not be properly positioned on each cell due to the $\lambda$-rules that generated errors due to the distance active contacts, poly contacts, and polysilicon routing that were positioned to minimize area.  To remove all layout errors, the PSelect was placed on the top rows of cells and the NSelect was placed on the bottom cells, but this is not correct since the PSelect needs to be on the top part of each cell (around the PMOSs) and the NSelect on the bottom part of each cell (around the NMOs.  In other words, each cell requires both a PSelect and NSelect as in Figure \ref{fig:cell}.  
\begin{lstlisting}
 Cpar1 1 0 C=1.82822p
Cpar2 2 0 C=13.0335f
Cpar3 3 0 C=12.971f
Cpar4 4 0 C=13.0335f
Cpar5 5 0 C=13.379f
Cpar6 6 0 C=24.049f
Cpar7 7 0 C=24.29725f
Cpar8 10 0 C=17.01275f
Cpar9 14 0 C=23.6905f
Cpar10 15 0 C=23.6905f
Cpar11 16 0 C=19.286f
Cpar12 17 0 C=18.464f
Cpar13 18 0 C=22.892f
Cpar14 19 0 C=23.628f
Cpar15 20 0 C=23.6905f
Cpar16 21 0 C=23.6905f
Cpar17 24 0 C=13.984f
Cpar18 25 0 C=13.984f

M19 ? 1 ? 1 NMOS L=0u W=0u 
* M19 DRAIN GATE SOURCE BULK (298.5 3 300.5 10) 
M20 ? 1 ? 1 NMOS L=0u W=0u 
* M20 DRAIN GATE SOURCE BULK (298.5 29 300.5 34.5) 
M21 ? 1 ? 1 NMOS L=0u W=0u 
* M21 DRAIN GATE SOURCE BULK (298.5 20.5 300.5 26) 
M22 1 20 14 1 NMOS L=2u W=6.5u AD=1.43025n PD=610u AS=138.5p PS=70u 
* M22 DRAIN GATE SOURCE BULK (133 -28 135 -21.5) 
M23 1 20 14 1 NMOS L=2u W=10.5u AD=1.43025n PD=610u AS=138.5p PS=70u 
* M23 DRAIN GATE SOURCE BULK (133 -58.5 135 -48) 
M24 2 1 19 1 NMOS L=2.5u W=11.5u AD=63.25p PD=34u AS=138.5p PS=70u 
* M24 DRAIN GATE SOURCE BULK (208.5 -33 211 -21.5) 
M25 18 1 25 1 NMOS L=2.5u W=11.5u AD=132.75p PD=69u AS=109.25p PS=42u 
* M25 DRAIN GATE SOURCE BULK (177.5 -33 180 -21.5) 
M26 ? 1 ? 1 NMOS L=0u W=0u 
* M26 DRAIN GATE SOURCE BULK (93 4.5 95 11) 
M27 ? 1 ? 1 NMOS L=0u W=0u 
* M27 DRAIN GATE SOURCE BULK (93 31 95 41.5) 
M28 ? 1 ? 1 NMOS L=0u W=0u 
* M28 DRAIN GATE SOURCE BULK (80 4.5 82 11) 
M29 ? 1 ? 1 NMOS L=0u W=0u 
* M29 DRAIN GATE SOURCE BULK (80 31 82 41.5) 
M30 ? 1 ? 1 NMOS L=0u W=0u 
* M30 DRAIN GATE SOURCE BULK (133.5 4.5 135.5 11) 
M31 ? 1 ? 1 NMOS L=0u W=0u 
* M31 DRAIN GATE SOURCE BULK (133.5 31 135.5 41.5) 
M32 ? 1 ? 1 NMOS L=0u W=0u 
* M32 DRAIN GATE SOURCE BULK (200 4.5 202 11) 
M33 ? 1 ? 1 NMOS L=0u W=0u 
* M33 DRAIN GATE SOURCE BULK (200 31 202 41.5) 
M34 ? 1 ? 1 NMOS L=0u W=0u 
* M34 DRAIN GATE SOURCE BULK (187 4.5 189 11) 
M35 ? 1 ? 1 NMOS L=0u W=0u 
* M35 DRAIN GATE SOURCE BULK (187 31 189 41.5) 
M36 ? 1 ? 1 NMOS L=0u W=0u 
* M36 DRAIN GATE SOURCE BULK (236 1 247.5 3) 
M37 ? 1 ? 1 NMOS L=0u W=0u 
* M37 DRAIN GATE SOURCE BULK (273 8 275 14.5) 
M38 10 1 1 1 NMOS L=2u W=6.5u AD=110.5p PD=60u AS=1.43025n PS=610u 
* M38 DRAIN GATE SOURCE BULK (272.5 -31.5 274.5 -25) 
M39 10 1 1 1 NMOS L=2u W=10.5u AD=110.5p PD=60u AS=1.43025n PS=610u 
* M39 DRAIN GATE SOURCE BULK (272.5 -50.5 274.5 -40) 
M40 ? 1 ? 1 NMOS L=0u W=0u 
* M40 DRAIN GATE SOURCE BULK (273 23 275 33.5) 
M41 ? 22 ? 1 NMOS L=0u W=0u 
* M41 DRAIN GATE SOURCE BULK (254 9 256 14.5) 
M42 ? 22 ? 1 NMOS L=0u W=0u 
* M42 DRAIN GATE SOURCE BULK (254 23 256 33.5) 
M43 ? 1 ? 1 NMOS L=0u W=0u 
* M43 DRAIN GATE SOURCE BULK (155.5 4.5 158 16) 
M44 ? 1 ? 1 NMOS L=0u W=0u 
* M44 DRAIN GATE SOURCE BULK (124.5 4.5 126.5 16) 
M45 1 8 16 1 NMOS L=2u W=5.5u AD=1.43025n PD=610u AS=112p PS=60u 
* M45 DRAIN GATE SOURCE BULK (226.5 -31.5 228.5 -26) 
M46 1 8 16 1 NMOS L=2u W=10.5u AD=1.43025n PD=610u AS=112p PS=60u 
* M46 DRAIN GATE SOURCE BULK (226.5 -50.5 228.5 -40) 
M47 1 9 1 1 NMOS L=2u W=11.5u AD=1.43025n PD=610u AS=1.43025n PS=610u 
* M47 DRAIN GATE SOURCE BULK (235.5 -58 247 -56) 
M48 6 1 7 1 NMOS L=2u W=5.5u AD=158p PD=92u AS=147p PS=77u 
* M48 DRAIN GATE SOURCE BULK (298 -51 300 -45.5) 
M49 6 1 5 1 NMOS L=2u W=5.5u AD=158p PD=92u AS=68p PS=47u 
* M49 DRAIN GATE SOURCE BULK (298 -42.5 300 -37) 
M50 5 1 7 1 NMOS L=2u W=5u AD=68p PD=47u AS=147p PS=77u 
* M50 DRAIN GATE SOURCE BULK (298 -26.5 300 -21.5) 
M51 17 11 1 1 NMOS L=2u W=5.5u AD=106.75p PD=59u AS=1.43025n PS=610u 
* M51 DRAIN GATE SOURCE BULK (253.5 -31.5 255.5 -26) 
M52 17 11 1 1 NMOS L=2u W=10.5u AD=106.75p PD=59u AS=1.43025n PS=610u 
* M52 DRAIN GATE SOURCE BULK (253.5 -50.5 255.5 -40) 
M53 4 1 21 1 NMOS L=2.5u W=11.5u AD=63.25p PD=34u AS=138.5p PS=70u 
* M53 DRAIN GATE SOURCE BULK (101.5 -33 104 -21.5) 
M54 15 1 1 1 NMOS L=2u W=11.5u AD=138.5p PD=70u AS=1.43025n PS=610u 
* M54 DRAIN GATE SOURCE BULK (70.5 -33 72.5 -21.5) 
M55 3 1 20 1 NMOS L=2.5u W=11.5u AD=63.25p PD=34u AS=138.5p PS=70u 
* M55 DRAIN GATE SOURCE BULK (155 -33 157.5 -21.5) 
M56 14 1 24 1 NMOS L=2u W=11.5u AD=138.5p PD=70u AS=109.25p PS=42u 
* M56 DRAIN GATE SOURCE BULK (124 -33 126 -21.5) 
M57 ? 1 ? 1 NMOS L=0u W=0u 
* M57 DRAIN GATE SOURCE BULK (311 20 316.5 22) 
M58 6 1 12 1 NMOS L=2u W=5.5u AD=158p PD=92u AS=33p PS=23u 
* M58 DRAIN GATE SOURCE BULK (310.5 -39 316 -37) 
M59 ? 1 ? 1 NMOS L=0u W=0u 
* M59 DRAIN GATE SOURCE BULK (236 38.5 247.5 40.5) 
M60 13 1 6 1 NMOS L=2u W=5.5u AD=33p PD=23u AS=158p PS=92u 
* M60 DRAIN GATE SOURCE BULK (310.5 -48 316 -46) 
M61 21 15 1 1 NMOS L=2u W=6.5u AD=138.5p PD=70u AS=1.43025n PS=610u 
* M61 DRAIN GATE SOURCE BULK (92.5 -28 94.5 -21.5) 
M62 21 15 1 1 NMOS L=2u W=10.5u AD=138.5p PD=70u AS=1.43025n PS=610u 
* M62 DRAIN GATE SOURCE BULK (92.5 -58.5 94.5 -48) 
M63 1 21 15 1 NMOS L=2u W=6.5u AD=1.43025n PD=610u AS=138.5p PS=70u 
* M63 DRAIN GATE SOURCE BULK (79.5 -28 81.5 -21.5) 
M64 1 21 15 1 NMOS L=2u W=10.5u AD=1.43025n PD=610u AS=138.5p PS=70u 
* M64 DRAIN GATE SOURCE BULK (79.5 -58.5 81.5 -48) 
M65 20 14 1 1 NMOS L=2u W=6.5u AD=138.5p PD=70u AS=1.43025n PS=610u 
* M65 DRAIN GATE SOURCE BULK (146 -28 148 -21.5) 
M66 20 14 1 1 NMOS L=2u W=10.5u AD=138.5p PD=70u AS=1.43025n PS=610u 
* M66 DRAIN GATE SOURCE BULK (146 -58.5 148 -48) 
M67 19 18 1 1 NMOS L=2u W=6.5u AD=138.5p PD=70u AS=1.43025n PS=610u 
* M67 DRAIN GATE SOURCE BULK (199.5 -28 201.5 -21.5) 
M68 19 18 1 1 NMOS L=2u W=10.5u AD=138.5p PD=70u AS=1.43025n PS=610u 
* M68 DRAIN GATE SOURCE BULK (199.5 -58.5 201.5 -48) 
M69 1 19 18 1 NMOS L=2u W=6.5u AD=1.43025n PD=610u AS=132.75p PS=69u 
* M69 DRAIN GATE SOURCE BULK (186.5 -28 188.5 -21.5) 
M70 1 19 18 1 NMOS L=2u W=10.5u AD=1.43025n PD=610u AS=132.75p PS=69u 
* M70 DRAIN GATE SOURCE BULK (186.5 -58.5 188.5 -48) 
M71 ? 1 ? 1 NMOS L=0u W=0u 
* M71 DRAIN GATE SOURCE BULK (311 29 316.5 31) 
M72 ? 1 ? 1 NMOS L=0u W=0u 
* M72 DRAIN GATE SOURCE BULK (146.5 4.5 148.5 11) 
M73 ? 1 ? 1 NMOS L=0u W=0u 
* M73 DRAIN GATE SOURCE BULK (146.5 31 148.5 41.5) 
M74 ? 23 ? 1 NMOS L=0u W=0u 
* M74 DRAIN GATE SOURCE BULK (227 9 229 14.5) 
M75 ? 23 ? 1 NMOS L=0u W=0u 
* M75 DRAIN GATE SOURCE BULK (227 23 229 33.5) 
M76 ? 1 ? 1 NMOS L=0u W=0u 
* M76 DRAIN GATE SOURCE BULK (209 4.5 211.5 16) 
M77 ? 1 ? 1 NMOS L=0u W=0u 
* M77 DRAIN GATE SOURCE BULK (178 4.5 180 16) 
M78 ? 1 ? 1 NMOS L=0u W=0u 
* M78 DRAIN GATE SOURCE BULK (102 4.5 104.5 16) 
M79 ? 1 ? 1 NMOS L=0u W=0u 
* M79 DRAIN GATE SOURCE BULK (71 4.5 73 16) 

* Total Nodes: 25
* Total Elements: 79
* Extract Elapsed Time: 0 seconds
.END   
\end{lstlisting}

\bibliographystyle{IEEEtran}
\bibliography{biblio.bib} 

\begin{thebibliography}{10}
\providecommand{\url}[1]{#1}
\csname url@samestyle\endcsname
\providecommand{\newblock}{\relax}
\providecommand{\bibinfo}[2]{#2}
\providecommand{\BIBentrySTDinterwordspacing}{\spaceskip=0pt\relax}
\providecommand{\BIBentryALTinterwordstretchfactor}{4}
\providecommand{\BIBentryALTinterwordspacing}{\spaceskip=\fontdimen2\font plus
\BIBentryALTinterwordstretchfactor\fontdimen3\font minus \fontdimen4\font\relax}
\providecommand{\BIBforeignlanguage}[2]{{%
\expandafter\ifx\csname l@#1\endcsname\relax
\typeout{** WARNING: IEEEtran.bst: No hyphenation pattern has been}%
\typeout{** loaded for the language `#1'. Using the pattern for}%
\typeout{** the default language instead.}%
\else
\language=\csname l@#1\endcsname
\fi
#2}}
\providecommand{\BIBdecl}{\relax}
\BIBdecl

\bibitem{Chang:2008}
L.~Chang, R.~Montoye, Y.~Nakamura, K.~Batson, R.~Eickemeyer, R.~Dennard, W.~Haenesch, and D.~Jamsek, ``An 8t sram for variability tolerance and low-voltage operation in high-performance caches,'' \emph{IEEE Journal of Solid State Circuits}, vol.~43, no.~4, pp. 956--963, 2008.

\bibitem{Lin:2008}
S.~Lin, Y.~Kim, and F.~Lombardi, ``A highly-stable nanometer memory for low-power design,'' \emph{Proceeding of IEEE International Workshop on Design and Test of Nano Devices, Circuits, and Systems}, pp. 17--20, 2008.

\bibitem{Liu:2008}
Z.~Liu and V.~Kursum, ``Characterization of a novel nine-transistor sram cell,'' \emph{IEEE Transactions on Very Large Scale Integration (VlSI) Systems}, vol.~16, no.~4, pp. 488--492, 2008.

\bibitem{Sivamangai:2011}
N.~Sivamangai and K.~Gunavathi, ``A low power sram cell with high read stability,'' \emph{ECTI Transactions on Electrical Eng., Electronics, and Communications}, vol.~9, no.~1, 2011.

\bibitem{Gul:2022}
W.~Gul, M.~Shams, and D.~Al-Khalili, ``Sram cell design challenges in modern deep sub-micron technologies: An overview,'' \emph{Micromachines}, vol.~13, 2022.

\bibitem{Mohagheghi:2022}
J.~Mohagheghi, B.~Ebrahimi, and P.~Torkzadeh, ``Single-ended 6t sram cell with low power/energy consumption and high stability,'' \emph{AUT Journal of Electrical Engineering}, vol.~54, pp. 343--360, 2022.

\bibitem{Qin:2000}
\BIBentryALTinterwordspacing
H.~Qin, ``Deep sub-micron sram design for ultra-low leakage standby operation,'' Ph.D. dissertation, University of California, Berkeley, 2000. [Online]. Available: \url{https://www2.eecs.berkeley.edu/Pubs/TechRpts/2007/EECS-2007-74.pdf}
\BIBentrySTDinterwordspacing

\bibitem{Yu:1997}
B.~Yu, W.~Lee, and C.~Hu, ``Modeling short-channel effects of cmosfet's taking account for channel-engineering, defect-enhanced-diffusion and gate-depletion,'' pp. 298--302, 1997.

\bibitem{Rabaey:1996}
J.~Rabaey, \emph{Digital Integrated Circuits: A Design Perspective}.\hskip 1em plus 0.5em minus 0.4em\relax Prentice Hall eleectronics and VLSI Series, 1996.

\bibitem{Hamouche:2012}
\BIBentryALTinterwordspacing
L.~Hamouche, ``Design of sram for cmos 32 nm,'' Ph.D. dissertation, INSA de Lyon, 2011. [Online]. Available: \url{https://theses.hal.science/tel-00715803/file/these.pdf}
\BIBentrySTDinterwordspacing

\bibitem{Lohstroh:1983}
J.~Lohstroh, E.~Seevinck, and J.~de~Groot, ``Worst-case static noise margin criteria for logic circuits and their mathematical equivalence,'' \emph{IEEE Jurnal of Solid State Circuits}, vol. SC-18, no.~6, pp. 803--807, 1983.

\end{thebibliography}

\end{document}